\documentclass[lettersize, journal]{IEEEtran}
\usepackage{algorithm}
\usepackage{algorithmic}
\usepackage{ifpdf}
\usepackage{cite}
\usepackage[cmex10]{amsmath}
\usepackage{array}
\usepackage{mdwmath}
\usepackage{mdwtab}
\usepackage{eqparbox}
\usepackage{amssymb} 
\usepackage{subfig}
\usepackage{amsfonts}
\usepackage{subcaption}
\usepackage{amsmath}
\usepackage{graphicx}
\usepackage{subfig}
\usepackage{float}  
\usepackage{bm}
\usepackage{subcaption}
\ifCLASSINFOpdf
\else
\fi
\begin{document}
\title{Performance Analysis of RIS-Aided High-Mobility Wireless Systems}

\author{Hanwen Hu, Jiancheng An,~\IEEEmembership{Member,~IEEE,} Lu Gan,~\IEEEmembership{Member,~IEEE,} and Chau Yuen, \emph{Fellow, IEEE}
        % <-this % stops a space
\thanks{H. Hu and L. Gan are with the School of Information and Communication Engineering, University of Electronic Science and Technology of China (UESTC), Chengdu, Sichuan, 611731, China. L. Gan is also with the Yibin Institute of UESTC, Yibin 644000, China (e-mail: hanwen\_hu\_uestc@outlook.com; ganlu@uestc.edu.cn).}% <-this % stops a space
\thanks {J. An and C. Yuen are with the School of Electrical and Electronics
 Engineering, Nanyang Technological University, Singapore 639798 (e-mail:
 jiancheng.an@ntu.edu.sg, chau.yuen@ntu.edu.sg).}}

\maketitle
\begin{abstract}
\textbf{Reconfigurable intelligent surface (RIS) technology holds immense potential for increasing the performance of wireless networks. Therefore, RIS is also regarded as one of the solutions to address communication challenges in high-mobility scenarios, such as Doppler shift and fast fading. This paper investigates a high-speed train (HST) multiple-input single-output (MISO) communication system aided by a RIS. We propose a block coordinate descent (BCD) algorithm to jointly optimize the RIS phase shifts and the transmit beamforming vectors to maximize the channel gain. Numerical results are provided to demonstrate that the proposed algorithm significantly enhances the system performance, achieving an average channel gain improvement of 15 dB compared to traditional schemes. Additionally, the introduction of RIS eliminates outage probability and improves key performance metrics such as achievable rate, channel capacity, and bit error rate (BER). These findings highlight the critical role of RIS in enhancing HST communication systems.}
\end{abstract}


\IEEEpeerreviewmaketitle
\section{Introduction}
With the rapid development of high-speed railways, ensuring reliable and high-quality wireless communication for high-speed trains (HSTs) has become a critical challenge\cite{ref0,ref1,ref2,ref3}. The increasing demand for seamless connectivity, real-time monitoring, and passenger entertainment systems requires robust communication systems that can handle the unique challenges posed by HST environments. Traditional communication systems often struggle to meet these demands due to severe Doppler shifts, fast fading, and high mobility, which degrade signal quality and limit system performance\cite{ref4}. As a result, there is a pressing need for innovative solutions to enhance the reliability and efficiency of HST communication systems.

Reconfigurable intelligent surfaces (RIS) have recently emerged as a transformative technology for wireless communication systems\cite{ref5,ref6,ref7,ref8}. Typically, a RIS consists of a large number of passive reflecting elements that can dynamically control the phase and amplitude of incident signals, thereby reshaping the propagation environment\cite{ref9,ref10,ref11}. By intelligently adjusting the reflection properties of RIS, it is possible to enhance signal strength, extend coverage, and mitigate interference. This makes RIS particularly well-suited for challenging environments such as HST communications, where traditional methods often fall short\cite{ref12,ref13,ref14}.

In this paper, we focus on the performance analysis of a RIS-assisted HST multiple-input single-output (MISO) communication system. The integration of RIS into HST communication systems presents several technical challenges, including the optimization of RIS phase shifts in real-time, the joint design of RIS and transmit beamforming, and the analysis of system performance under high-mobility conditions. Addressing these challenges requires novel algorithms and comprehensive performance evaluations.

The main contributions of this paper are as follows:
\begin{itemize}
    \item We propose a comprehensive system model for RIS-assisted HST MISO communications, considering both direct and cascaded links. The model accounts for the effects of high mobility, Doppler shifts, and fast fading.
    \item We formulate the channel gain maximization problem and propose an efficient block coordinate descent (BCD) algorithm to jointly optimize the RIS phase shifts and the transmit beamforming vectors. The algorithm addresses the challenges posed by high mobility and ensures robust performance under dynamic channel conditions.
    \item We conduct extensive numerical simulations to evaluate the performance of the proposed system. Key performance metrics, including channel gain, achievable rate, channel capacity, outage probability, and bit error rate (BER), are analyzed. The results demonstrate significant performance improvements compared to traditional schemes.
\end{itemize}

\textit{Notations}: Scalars are denoted by italic letters 
(\textit{a,b,...}). Vectors and matrices are denoted by bold-face lower-case ($\mathbf{a,b}$,...) and upper-case letters ($\mathbf{A,B}$,...). The $n$-th element of vector $\mathbf{a}$ is $a_n$. The element in the $n$-th row and $m$-th column of the matrix $\mathbf{A}$ is $A_{n,m}$. $j$ = $\sqrt{-1}$ denotes the imaginary unit. $\mathbb{C}^{M}$ denotes the set of $M \times 1$ complex vectors. $\mathbb{C}^{x \times y}$ and $\mathbb{R}^{x \times y}$ denote the spaces of $x \times y$ complex-valued and real-value matrices. $(\cdot)^\mathrm{T}$, $(\cdot)^\mathrm{H}$, $(\cdot)^*$, and $|\cdot |^2$ denote the transpose, conjugate transpose, conjugate operation, and 2-norm, respectively. For a complex-valued vector $\mathbf{x}$, $\operatorname{diag}(\mathbf{x})$ denotes a diagonal matrix with each diagonal element being the corresponding element in $\mathbf{x}$. $\otimes$  and $\odot$ denote the Kronecker product and Hadamard product, respectively. $\mathcal{CN}(\mathbf{a}, \mathbf{C})$ stands for the circularly-symmetric complex Gaussian distribution with mean vector $\mathbf{a}$ and covariance matrix $\mathbf{C}$.

\section{System Model and Problem Formulation}
\begin{figure}[t] 
\centering 
\includegraphics[width=0.3\textwidth]{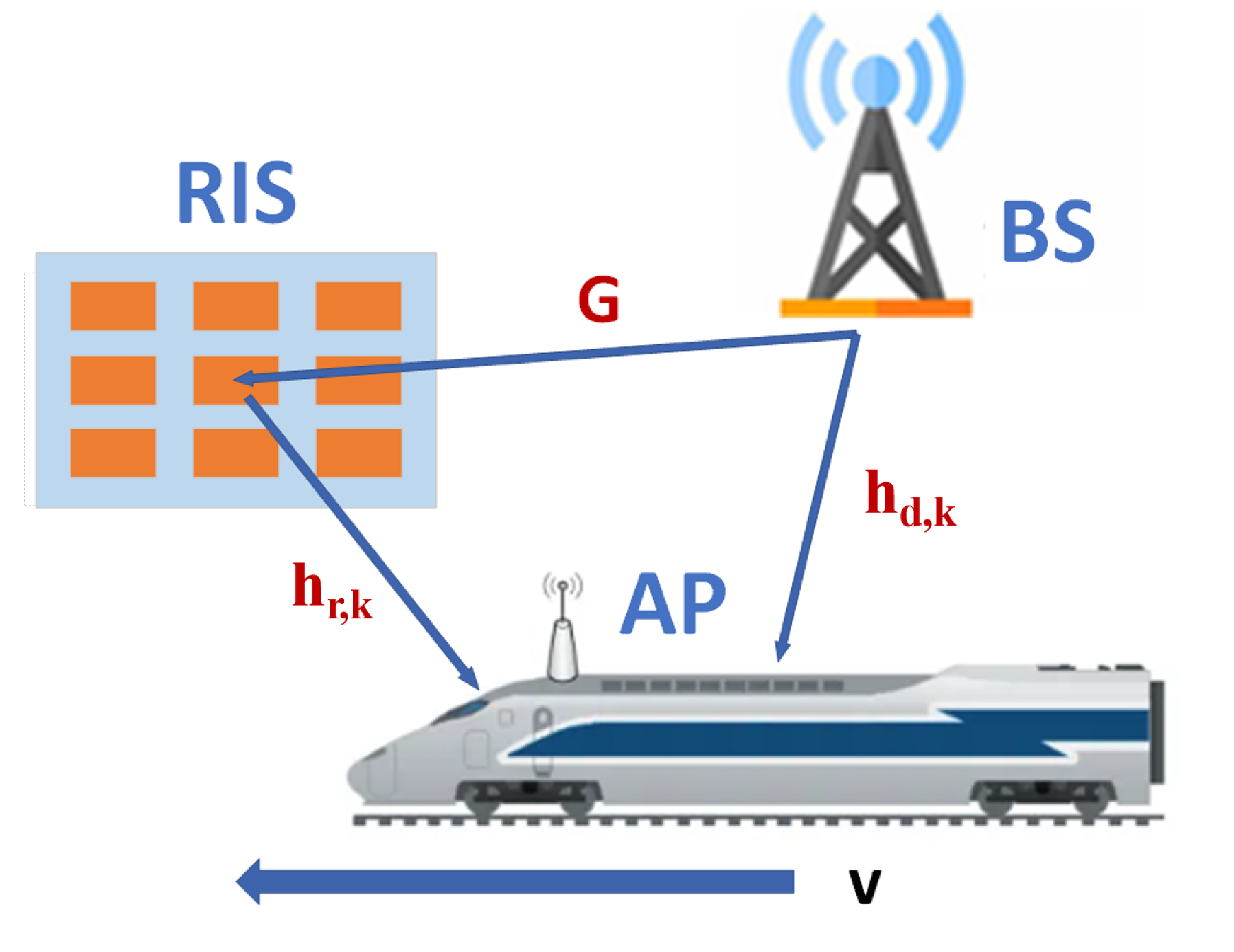} 
\captionsetup{font=small}
\caption{Illustration of a RIS-assisted MISO downlink system for HST communications.}
\label{Fig.main1}
\end{figure}
\subsection{HST MISO System Model}
As shown in Fig. 1, a downlink HST MISO communication system from the base station (BS) to a single-antenna access point (AP), assisted by a RIS composed of $N_I=N^2$ elements, is investigated. The BS is equipped with $M$ transmit antennas (TAs), arranged in a uniform linear array (ULA) with half-wavelength spacing. We assume that the train travels at a constant speed $v$, with a frame length of $ T_d $, each frame consists of $ K $ time slots of length $ T_c $, satisfying $ T_d = K \times T_c $. 

We consider two types of links: the BS-AP link (direct link) and the BS-RIS-AP link (cascaded link). The baseband equivalent channel $\mathbf{h}_{d,k} \in \mathbb{C}^{M} $ of the direct link in the $k$-th time slot follows a Rician distribution with factor $\kappa$ as
\begin{equation}
     \mathbf{h}_{d,k}= \sqrt{\frac{\kappa}{1 + \kappa}} \mathbf{h_d}_{\text{LOS},k} + \sqrt{\frac{1}{1 + \kappa}} \mathbf{h_d}_{\text{NLOS},k}, k = 1, 2, \dots, K.
\end{equation}

The cascaded link consists of the BS-RIS link $\mathbf{G} \in \mathbb{C}^{N_I \times M}$ and the RIS-AP $ \mathbf{h}_{r,k} \in \mathbb{C}^{ N_I} $ link in the $k$-th time slot. Since the location of the RIS is preplanned, we assume that the BS-RIS link is a line-of-sight (LOS) transmission. For a given azimuth arrival angle $\theta_1$ and an elevation arrival angle $\phi_1$, $\mathbf{G}$ is modeled as
\begin{equation}
    \mathbf{G} = \alpha \left(\mathbf{a}_y(\theta_1, \phi_1) \otimes\ \mathbf{a}_z(\phi_1)\right)\mathbf{a_{\text{BS}}}^H(\varphi),
    \end{equation}
where $\alpha$ and $\varphi$ denote the corresponding path loss and the transmit angle of TAs, respectively. $\mathbf{a}_y(\theta, \phi)\in \mathbb{C}^{N}$, $\mathbf{a}_z(\phi) \in \mathbb{C}^{N}$ and $\mathbf{a_{\text{BS}}}(\varphi) \in \mathbb{C}^{M}$ are defined by
\begin{equation}
\mathbf{a}_y(\theta, \phi) = 
\begin{bmatrix} 
1,e^{j\pi\sin{\theta}\cos{\phi}},\ldots,e^{j\pi(N-1)\sin{\theta}\cos{\phi}} 
\end{bmatrix}^T,
\end{equation}
\begin{equation}
\mathbf{a}_z(\phi) = 
\begin{bmatrix}
1,e^{j\pi\sin{\phi}},\ldots,e^{j\pi(N-1)\sin{\phi}} 
\end{bmatrix}^T,
\end{equation}
\begin{equation}
\mathbf{a_{\text{BS}}}(\varphi) = 
\begin{bmatrix}
1,e^{j\pi\sin{\varphi}},\ldots,e^{j\pi(M-1)\sin{\varphi}}
\end{bmatrix}^T.
\end{equation}

For the RIS-AP link, the channel $ \mathbf{h}_{r,n} $ in the $ k $-th time slot is modeled as a Rician channel, yielding
\begin{equation}
     \mathbf{h}_{r,k}= \sqrt{\frac{\kappa}{1 + \kappa}} \mathbf{h}_{\text{LOS},k} + \sqrt{\frac{1}{1 + \kappa}} \mathbf{h}_{\text{NLOS},k}, \quad k = 1, 2, \dots, K.
\end{equation}
The non-line-of-sight (NLOS) component follows a Rayleigh distribution, and its power spectral density is given by the Jakes spectrum. The LOS component $\mathbf{h}_{\text{LOS},k}\in \mathbb{C}^{ N_I}$, is only affected by the Doppler frequency shift, which is given by:
\begin{equation}
\mathbf{h}_{\text{LOS},k}(\theta_2, \phi_2) =\beta \left(\bm{a}_y(\theta_2, \phi_2) \otimes\ \bm{a}_z(\phi_2)\right) e^{j2\pi k f_dT_c}, 
\end{equation}
where $\beta$ denotes the corresponding path loss and $f_d$ denotes the Doppler frequency. $\theta_2$ and $\phi_2$ represent the azimuth and elevation departure angles, respectively.

Moreover, let $\boldsymbol{\Phi}_k = \text{diag}( e^{j\varphi_{1,k}}, \ldots, e^{j\varphi_{N_I,k}})$ denote the phase shift matrix of RIS in the $k$-th time slot, where $\varphi_{n,k} \in [0, 2\pi),n = 1, 2, \dots, N_I$ denotes the phase shift of the $n$-th RIS element. In this paper, we consider linear transmit precoding at the BS with one transmit beamforming vector in the $k$-th time slot. Hence, the complex baseband transmitted signal at the BS can be expressed as $\mathbf{x}_k = \mathbf{w}_k s$, where $s$ denotes the transmitted data and $\mathbf{w}_k \in \mathbb{C}^{M \times 1}$ is the corresponding beamforming vector. The signal received at the AP is then expressed as
\begin{equation}
    y_k = \left(\mathbf{h}_{r,k}^\mathrm{H} \boldsymbol{\Phi}_k  \mathbf{G} + \mathbf{h}_{d,k}^\mathrm{H}\right)\mathbf{w}_ks + n,
\end{equation}
where $n\sim \mathcal{CN}(0, \sigma^2)$ denotes the additive white Gaussian noise (AWGN). 
Therefore, we define the chanel gain of MISO system in the $k$-th time slot as $F(\mathbf{w}_k,\boldsymbol{\Phi}_k)$:
\begin{equation}
    F(\mathbf{w}_k,\boldsymbol{\Phi}_k)= |\left(\mathbf{h}_{r,k}^\mathrm{H} \boldsymbol{\Phi}_k  \mathbf{G} + \mathbf{h}_{d,k}^\mathrm{H}\right)\mathbf{w}_k|^2.
\end{equation}

\subsection{Problem Formulation}
In this paper, we aim to optimize $\mathbf{w}_k$ and $\boldsymbol{\Phi}_k$ to maximize the channel gain in the HST MISO communication model. The constraints of the problem include phase constraints, and the transmit power constraints. Accordingly, the problem is formulated as
\begin{align}
&\text{(P1)} :  \max_ {\mathbf{w}_k,\boldsymbol{\Phi}_k}  F(\mathbf{w}_k,\boldsymbol{\Phi}_k)=|\left(\mathbf{h}_{r,k}^\mathrm{H} \boldsymbol{\Phi}_k  \mathbf{G} + \mathbf{h}_{d,k}^\mathrm{H}\right)\mathbf{w}_k|^2 \tag{10a}\\
&\text{s.t.} \quad ||\bm{w}_k||^2 \leq P, \quad k = 1, \ldots, K.\tag{10b}\\
& \quad \quad0 \leq \varphi_{n,k} \leq 2\pi, \quad k = 1, \ldots, K,\quad n = 1, \ldots, N,\tag{10c}
\end{align}
where $P$ denotes the transmit power of the BS.

\section{Solution to the Channel Gain Maximizaion Problem}

In this section, we propose an effective block coordinate descent algorithm to tackle the channel gain maximization problem (P1) by jointly optimize the transmit beamforming vector $\mathbf{w}_k$ and the phase shift matrix $\boldsymbol{\Phi}_k$. Specifically,
the algorithm partitions the optimization variables into several blocks, optimizing one block per iteration while keeping the others fixed.

\subsection{Optimization with Respect to the Phase Shift $\boldsymbol{\Phi}_k$}
For a given transmit beamforming vector $\mathbf{w}_k$, the optimal RIS phase shift matrix $\boldsymbol{\Phi}_k$ should simultaneously eliminate the Doppler frequency shift of the direct channel in the $k$-th time slot and further constructively combine the LOS and NLOS channels. Therefore, optimization of the phase shift matrix $\boldsymbol{\Phi}_k$ mainly involves two steps.
\subsubsection{Eliminate the Doppler Frequency of the LOS Channel}To address the impact of the Doppler frequency shift caused by the LOS path of $\mathbf{h}_{r,k}$, we first perform the following transformation:
\setcounter{equation}{10}
\begin{equation}
\mathbf{h}_{\text{LOS},k}^\mathrm{H}\boldsymbol{\Phi}_k\mathbf{G}\mathbf{w}_k=\mathbf{v}_k^\mathrm{H}\mathbf{u}_k,
\end{equation} 
where $\mathbf{v}_k= [e^{j\varphi_{1,k}}, \ldots, e^{j\varphi_{N_I,k}}]^\mathrm{H}$. And $\mathbf{u}_k$ is derived as
\begin{align}
    \mathbf{u}_k&=\text{diag}(\mathbf{h}_{\text{LOS},k}^\mathrm{H})\mathbf{G}\mathbf{w}_k\notag\\
    &=\beta e^{j2\pi k f_dT_c}\text{diag}\left(\bm{a}_y(\theta_2, \phi_2) \otimes\ \bm{a}_z(\phi_2)\right)\mathbf{G}\mathbf{w}_k.
\end{align}
Thereby, in the $k$-th time slot, the RIS phase is adjusted as:
\begin{equation}
    \mathbf{v}^\mathrm{H}_{k,(1)} = e^{-j 2 \pi  kf_d T_c} \left(\text{diag}\left(\bm{a}_y(\theta_2, \phi_2) \otimes\ \bm{a}_z(\phi_2)\right)\mathbf{G}\mathbf{w}_k\right)^\mathrm{H}.
\end{equation}
This adjustment eliminates the Doppler frequency shift introduced in $\mathbf{h}_{\text{LOS},k}$ and enables RIS beamforming, thereby enhancing the LOS path gain in the cascaded channel. After RIS beamforming, at the BS side, the LOS path gain in the cascaded channel becomes significantly larger and dominates among all channel components. This improves the fast-fading effects of the equivalent composite channel.
\subsubsection{Align the Direct Channel and the Cascaded Channel}
By substituting (13) into (6) and (11), we have the cascaded channel after beamforming represented as
\begin{equation}
    h_{k,(1)}=\sqrt{\frac{\kappa}{1 + \kappa}} \mathbf{v}^\mathrm{H}_{k,(1)}\mathbf{u}_k + \sqrt{\frac{1}{1 + \kappa}} \mathbf{h}_{\text{NLOS},k}\boldsymbol{\Phi}_k\mathbf{G}\mathbf{w}_k.
\end{equation}
Thus the baseband equivalent channel spanning from BS to AP after beamforming is $h_{k,(1)}+\mathbf{h}_{d,k}^\mathrm{H}\mathbf{w}_k$. Since adding an arbitrary phase rotation to the beamforming vector does not affect the beamforming gain, we define $e^{j \boldsymbol{\epsilon}_k}$ as the adjusted phase obtained in the second step, yielding the channel gain:
\begin{equation}
    F(\mathbf{w}_k,\boldsymbol{\Phi}_k)=|h_{k,(1)}e^{j \boldsymbol{\epsilon}_k}+\mathbf{h}_{d,k}^\mathrm{H}\mathbf{w}_k|^2.
\end{equation}
The above equation admits a closed-form solution by exploiting the special structure of its objective function. Specifically, we have the following inequality:
\begin{equation}
     |h_{k,(1)}e^{j \boldsymbol{\epsilon}_k}+\mathbf{h}_{d,k}^\mathrm{H}\mathbf{w}_k| \overset{(a)}{\leq} |h_{k,(1)}e^{j \boldsymbol{\epsilon}_k}|+|\mathbf{h}_{d,k}^\mathrm{H}\mathbf{w}_k|,
\end{equation}
where (a) is due to the triangle inequality. So the optimal solution of $\boldsymbol{\epsilon}_k$ is obtained by
\begin{equation}
 e^{j\boldsymbol{\epsilon}_k^*} = e^{-j( \arg(h_{k,(1)}-\arg(\mathbf{h}_{d,k}^\mathrm{H}\mathbf{w}_k))}.
\end{equation}

Through combining the above two steps, we can obtain the optimal phase shift matrix $\boldsymbol{\Phi}_k^*=\text{diag}(\mathbf{v}_{k,(1)}e^{j\boldsymbol{\epsilon}_k^*})$
By doing so, the RIS-reflected signal received at the AP can coherently combine with the signal transmitted from the BS, thereby improving the overall received channel gain at the AP.
\subsection{Optimization with Respect to the Transmit Beamforming $\mathbf{w}_k$}
Given the RIS phase shift $\boldsymbol{\Phi_k}$, the optimal transmit beamforming vector $\mathbf{w}_k$ is the well-known maximal ratio transmission (MRT) beamforming along the same direction as the channel, i.e.,
\begin{equation}
  \mathbf{w}_k^*=\sqrt{P}\frac{(\mathbf{h}_{r,k}^\mathrm{H} \boldsymbol{\Phi}_k  \mathbf{G} + \mathbf{h}_{d,k}^\mathrm{H})^\mathrm{H}}{\left|\left|(\mathbf{h}_{r,k}^\mathrm{H} \boldsymbol{\Phi}_k  \mathbf{G} + \mathbf{h}_{d,k}^\mathrm{H})\right|\right|}.
\end{equation}

Based on the above analysis and discussions, we have provided the detailed procedure of the proposed BCD optimization algorithm, as outlined in Algorithm 1.
\begin{algorithm}[t]
\caption{The Proposed Block Coordinate Descent Algorithm for RIS-aided HST MISO System }
\begin{algorithmic}[1]
\STATE Initialize the transmit beamforming vector $\mathbf{w}_k^{(0)}$ and set the iteration counter to $i$ = 0
\REPEAT
    \STATE Optimize $\boldsymbol{\Phi_k}$ with given $\mathbf{w}_k^{(i)}$, and denote the optimal beamforming as $\boldsymbol{\Phi_k}^{(i+1)}$
    \STATE Optimize $\mathbf{w}_k$ with given $\boldsymbol{\Phi_k}^{(i+1)}$ and denote the optimal transmit beamforming as $\mathbf{w}_k^{(i+1)}$
    \STATE Increase the iteration counter by $i \leftarrow i + 1$.
\UNTIL{The objective function coverages or $i$ exceeds the maximum tolerable number of iterations.}
\end{algorithmic}
\end{algorithm}

\begin{figure*}[!t]
\centering
\subfloat[]{\includegraphics[width=6.2cm]{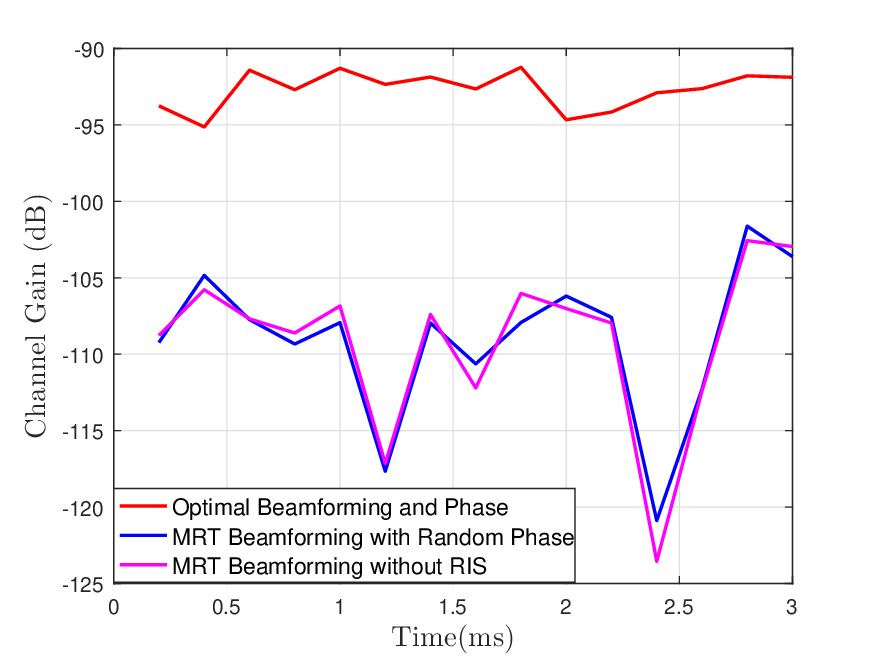}}
\subfloat[]{\includegraphics[width=6.2cm]{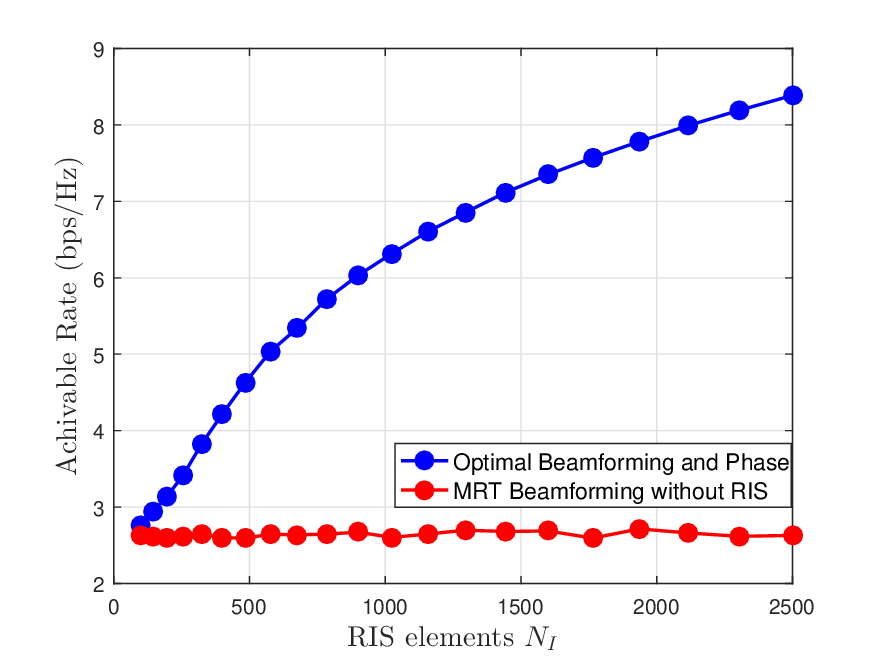}}
\subfloat[]{\includegraphics[width=6.2cm]{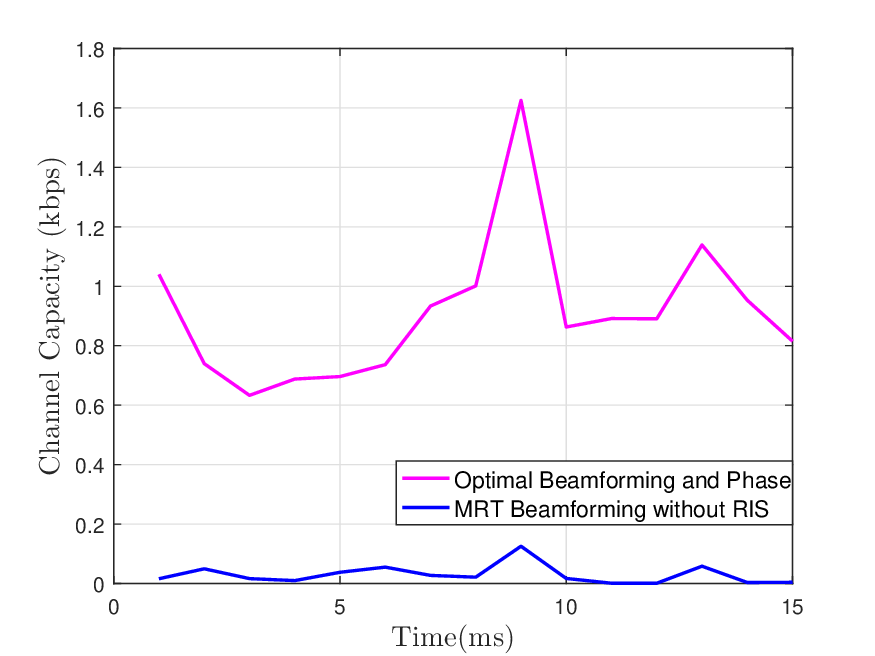}}
\caption{(a) Channel Gain versus time slots; (b) Achievable rate versus RIS elements $N_I$; (c) Channel capacity versus time slots.}
\label{fig_2}
\vspace{-0.5cm}
\end{figure*}

\section{Main Performance Analysis Metrics}
In this section, we present several main performance analysis metrics related to channel gain for the RIS-assisted HST MISO communication system and evaluate the impact of the introduction of RIS on the performance.
\subsection{Achivable Rate}
In the $k$-th time slot, after adjusting the RIS to the optimal phase $\boldsymbol{\Phi}_k^*$, the achievable rate is given by:
\begin{equation}
    R = \frac{1}{K} \sum_{k=1}^{K} \log\left(1 + \frac{|\hat{h}_k|^2}{\Gamma \sigma^2}\right),
\end{equation}
where $ \hat{h}_k = h_{k,(1)}e^{j \boldsymbol{\epsilon}_k^*}+\mathbf{h}_{d,k}^\mathrm{H}\mathbf{w}_k $ represents the maximum channel gain in the $k$-th time slot after the RIS is adjusted to the optimal phase, $ \Gamma \geq 1 $ denotes the gap to the actual channel capacity, which is determined by the modulation and coding scheme (MCS) used.
\subsection{Channel Capacity}
Channel capacity represents the maximum transmission rate of the RIS-assisted HST MISO system, expressed as
\begin{equation}
    C =  \frac{1}{K}\sum_{k=1}^{K}B \log_2 \left(1 + \frac{|\hat{h}_k|^2}{\sigma^2}\right),
\end{equation}
where $B$ is the bandwidth of the channel. This formula quantifies the theoretical upper limit of the data rate that can be achieved in the RIS-assisted HST MISO system under optimal RIS phase shift and transmit beamforming.
\subsection{Outage Probability }
Outage Probability is a key performance metric in wireless communication systems, used to measure the probability that the system fails to meet a specific quality of service (QoS) requirement under given channel conditions. Specifically, the outage probability is defined as the probability that the received signal-to-noise ratio (SNR) $\gamma$ falls below a certain threshold $ \gamma_{\text{th}} $. Mathematically, the outage probability can be expressed as:
\begin{equation}
    P_{\text{out},k} = P\left(\gamma_k < \gamma_{\text{th}}\right)=P\left(|\hat{h}_k|^2<\sigma^2\gamma_{\text{th}}\right).
\end{equation}
The probability distribution of $\hat{h}_k$ follows a complex-valued Gaussian distribution with mean $ \mu_{h,k} $ and variance $ \sigma_{h,k}^2 $, namely, $\hat{h}_k \sim \mathcal{CN}(\mu_{h,k},\sigma_{h,k}^2) $, and we have:
\begin{align}
\mu_{h,k} = \rho\mathbf{h}_{d,k}^\mathrm{H}\mathbf{w}_k + \rho^2 h_{k,(1)}e^{j \boldsymbol{\epsilon}_k^*},
\end{align}
\begin{equation}
\sigma_{h,k}^2 = \varrho^2|\mathbf{h}_{d,k}^\mathrm{H}\mathbf{w}_k|^2+ \varrho^4|h_{k,(1)}e^{j \boldsymbol{\epsilon}_k^*}|^2,
\end{equation}
where $ \rho =\sqrt{\kappa/1 + \kappa}$ and $\varrho=\sqrt{1/1 + \kappa} $.

Hence, the coverage probability  $P_{\text{out},k}$ follows a non-central chi-square distribution, i.e., $ \chi^2(\eta, \zeta_k) $, with the degree of freedom $ \eta = 2 $ and the non-centrality parameter $ \zeta_k = |\mu_{h,k}|^2/\sigma_{h,k}^2 $. With the corresponding cumulative distribution function (CDF), $P_{\text{out},k}$ is given by:
\begin{equation}
P_{\text{out},k} = 1 - Q_1\left( \sqrt{\zeta_k}, \sqrt{\gamma_{0,k}} \right),
\end{equation}
where $ \gamma_{0,k} = \sigma^2\gamma_{\text{th}}/\sigma_{h,k}^2 $, and $ Q_m(a, b) $ is the Marcum Q-function. From the above discussion, the introduction of RIS can alter the channel characteristics, thereby optimizing the distribution of the outage probability.

\begin{figure}[!t]
\centering
\subfloat[]{\includegraphics[width=4.6cm]{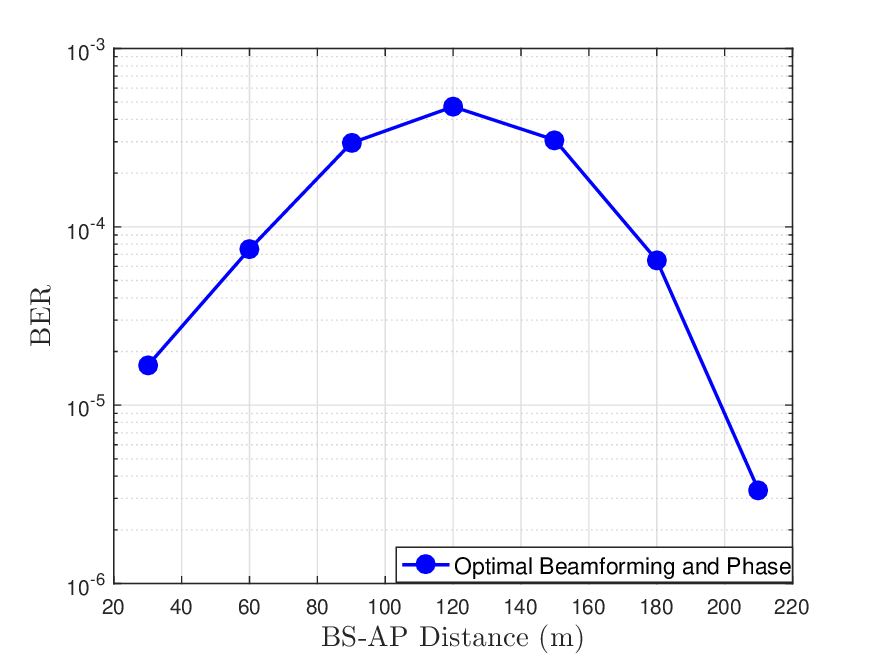}}
\subfloat[]{\includegraphics[width=4.6cm]{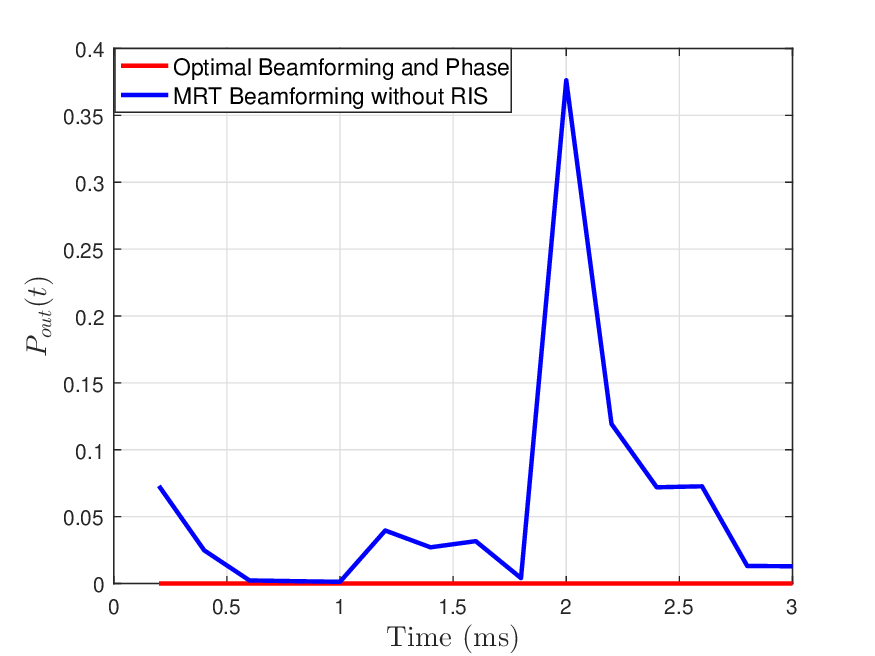}}
\caption{(a) BER versus AP position; (b) Outrage probability versus time slots.}
\label{fig_2}
\vspace{-0.5cm}
\end{figure}
\subsection{Bit Error Rate (BER)}
BER $P_b$ is a key metric in digital communication systems, representing the ratio of erroneous bits to the total number of transmitted bits. It reflects the impact of noise, interference, and channel fading on signal transmission. A lower BER indicates higher system reliability. For BPSK modulation, the BER is given by
\begin{equation}
    P_b = Q\left(\sqrt{\frac{2E_b}{\sigma^2}}\right),
\end{equation}
where $E_b$ is the energy per bit.
\begin{table}[t!]
\centering
\caption{Simulation Parameters}
\renewcommand{\arraystretch}{1.3} 
\begin{tabular}{|p{5cm}|p{3cm}|}
\hline
\textbf{Simulation Parameters} & \textbf{Value} \\ \hline
High-speed train speed \( v \) & 360 km/h \\ \hline
Carrier frequency \( f_c \) & 5 GHz \\ \hline
Bandwidth \( B \) & 100 kHz \\ \hline
Noise power \( \sigma^2 \) & -90 dBm \\ \hline
Transmit power \( P_i \) & 40 dBm \\ \hline
Number of the TAs \(M\) & 2 \\ \hline
Frame length \( T_d \) & 3 ms \\ \hline
Reference distance (1 m) path loss \( C_0 \) & 30 dB \\ \hline
Rician factor \( \kappa \) & 3 \\ \hline
Path loss exponent for BS–AP link $\ell_{\text{BA}}$\textbf{}& 3.8 \\ \hline
Path loss exponent for BS–RIS link $\ell_{\text{BR}}$& 2.2 \\ \hline
Path loss exponent for RIS–AP link$\ell_{\text{RA}}$ & 2.8 \\ \hline
Doppler power spectrum & Jakes \\ \hline
Coordinates of the BS & [0,0,30] \\ \hline
Coordinates of the RIS & [0,300,30] \\ \hline
\end{tabular}
\end{table}

\section{Simulation Result}

In this section, we provide numerical results to validate our proposed design for RIS-aided HST MISO communication. The simulation parameters are set as listed in Table I. For comparison, another three schemes are considered:
\begin{itemize}
    \item \textbf{Proposed algorithm}: The phase shift of RIS and the transmit beamforming are both optimized according to the proposed algorithm.
    \item \textbf{MRT Beamforming with Random Phase}: This scheme randomly selects a value of phase $\mathbf{v_k}$ for RIS elements and keeps unchanged, while MRT beamforming is utilized at the BS.
    \item \textbf{MRT Beamforming without RIS}: This scheme sets the phase shift to 0 and only MRT beamforming works.
\end{itemize}

Specifically, we consider a typical HST MISO downlink communication system assisted by a RIS with $N_I=N^2$ elements. The distance-dependent path loss $\alpha \sim  \mathcal{CN}(0, \rho_R^2)$ and  $\beta\sim  \mathcal{CN}(0, \rho_K^2)$ follow the free space path loss model with
$\rho_R^2 = C_0 \left( d_{\text{BR}}/d_0 \right)^{-\ell_{\text{BR}}}$ and $\rho_K^2 = C_0 \left( d_{\text{RA}}/d_0 \right)^{\ell_{\text{RA}}}$, where $C_0$ is the path loss at the reference distance $d_0$ = 1 m. $d_{\text{BR}}$ and $d_{\text{RA}}$ are the dictance of BS-RIS link and RIS-AP link, respectively. The angles $\phi_1, \phi_2, \theta_1,\theta_2, \varphi$ are continuous and uniformly distributed over  $[-\pi/2, \pi/2]$.  Unless otherwise specified, the coordinate of the AP is set to $[20,300,0]$.

In Fig. 2 (a), we demonstrate the effectiveness of the proposed BCD-based channel gain optimization algorithm under the condition of $N_I=1600$. The optimal phase obtained by the algorithm, combined with beamforming, brings an average channel gain improvement of 15 dB compared to the other two schemes. Additionally, it can be observed that the impact of random phases on channel gain is negligible, further proving the importance of optimizing the phase shift of RIS.

Fig. 2 (b) illustrates the relationship between the achievable rate and the number of the RIS elements $N_I$. We employ BPSK modulation with a modulation efficiency factor of 0.8. It can be observed that as $N_I$ increases, the achievable rate of the RIS-assisted HST system also increases. As expected, the achievable rate of the system without RIS assistance shows no dependence on $N_I$.

We have also investigated the channel capacity under different time slots, as shown in Fig. 2 (c). The number of RIS elements is set as $N_I=1600$. It is observed that the introduction of RIS significantly enhances the channel capacity of the system at all times compared to the traditional HST MISO system. The maximum capacity gap reaches approximately 1.5 kbps.

In Fig. 3 (a), we vary the position of the AP to investigate the relationship between the AP's location and the BER. We fix the number of RIS elements to \( N_I = 1600 \), and the BS transmits BPSK signals. It can be observed that when the RIS is adjusted to the optimal phase, as the train moves away from the BS and gradually approaches the RIS, the BER initially increases and then decreases. This is because when the train is close to the BS, the path loss between the AP and the BS is smaller, leading to a higher received SNR and thus a lower BER. When the train approaches the RIS, the AP receives weaker signals from the BS, but since the AP is closer to the RIS, it can receive stronger reflected signals from the RIS. This ensures that the received SNR remains at a relatively high level, keeping the BER low. When the train is in the middle position (equidistant from both the BS and the RIS), the received SNR at the AP is relatively lower, resulting in a higher BER.

Fig. 3 (b) demonstrates the outage probability of the RIS-assisted HST MISO system over the entire time period. Here, we set the number of RIS elements to \( N_I = 1600 \) and the SNR threshold to \( \gamma_{\text{th}} = 10 \, \text{dB} \). It can be observed that the system without RIS assistance experiences a certain probability of outage, while the introduction of RIS eliminates this possibility of outage. This further proves the performance enhancement of RIS for the HST system.

\section{Conclusion}
In this paper, we have investigated the performance of a RIS-assisted HST MISO communication system. By jointly optimizing the RIS phase shifts and the transmit beamforming vectors, we significantly improved the channel gain, achieving an average enhancement of 15 dB compared to traditional schemes. The proposed BCD algorithm effectively addressed the challenges posed by high mobility and Doppler effects. Numerical results confirmed that the introduction of RIS not only improves achievable rate and channel capacity but also eliminates outage probability and reduces BER. These findings underscore the importance of RIS in future wireless communication systems, particularly in high-mobility scenarios such as HSTs. Future work will explore more scenarios for RIS-assisted HST systems, such as multiple input multiple output (MIMO) systems, to further optimize system performance.
\bibliographystyle{IEEEtran}
\bibliography{wireless}

\begin{thebibliography}{10}
\providecommand{\url}[1]{#1}
\csname url@samestyle\endcsname
\providecommand{\newblock}{\relax}
\providecommand{\bibinfo}[2]{#2}
\providecommand{\BIBentrySTDinterwordspacing}{\spaceskip=0pt\relax}
\providecommand{\BIBentryALTinterwordstretchfactor}{4}
\providecommand{\BIBentryALTinterwordspacing}{\spaceskip=\fontdimen2\font plus
\BIBentryALTinterwordstretchfactor\fontdimen3\font minus \fontdimen4\font\relax}
\providecommand{\BIBforeignlanguage}[2]{{%
\expandafter\ifx\csname l@#1\endcsname\relax
\typeout{** WARNING: IEEEtran.bst: No hyphenation pattern has been}%
\typeout{** loaded for the language `#1'. Using the pattern for}%
\typeout{** the default language instead.}%
\else
\language=\csname l@#1\endcsname
\fi
#2}}
\providecommand{\BIBdecl}{\relax}
\BIBdecl

\bibitem{ref0}
B.~Ai \emph{et~al.}, ``Future railway services-oriented mobile communications network,'' \emph{IEEE Commun. Mag.}, vol.~53, no.~10, pp. 78--85, Oct 2015.

\bibitem{ref1}
R.~He, B.~Ai, G.~Wang, K.~Guan, Z.~Zhong, A.~F. Molisch, C.~B. Rodriguez, and C.~P. Oestges, ``High-speed railway communications: From {GSM-R} to {LTE-R},'' \emph{IEEE Veh. Technol. Mag.}, vol.~11, no.~3, pp. 49--58, Sep 2016.

\bibitem{ref2}
J.~Zhang, H.~Du, P.~Zhang, J.~Cheng, and L.~Yang, ``Performance analysis of {5G} mobile relay systems for high-speed trains,'' \emph{IEEE J. Sel. Areas Commun.}, vol.~38, no.~12, pp. 2760--2772, Dec 2020.

\bibitem{ref3}
R.~He \emph{et~al.}, ``{5G} for railways: The next generation railway dedicated communications,'' \emph{IEEE Commun. Mag.}, vol.~60, no.~12, pp. 130--136, Dec 2022.

\bibitem{ref4}
B.~Ai \emph{et~al.}, ``Challenges toward wireless communications for high-speed railway,'' \emph{IEEE Trans. Intell. Transp. Syst.}, vol.~15, no.~5, pp. 2143--2158, Oct 2014.

\bibitem{ref5}
M.~Di~Renzo, A.~Zappone, M.~Debbah, M.-S. Alouini, C.~Yuen, J.~de~Rosny, and S.~Tretyakov, ``Smart radio environments empowered by reconfigurable intelligent surfaces: How it works, state of research, and the road ahead,'' \emph{IEEE J. Sel. Areas Commun.}, vol.~38, no.~11, pp. 2450--2525, Nov. 2020.

\bibitem{ref6}
Y.~Liu, X.~Liu, X.~Mu, T.~Hou, J.~Xu, M.~Di~Renzo, and N.~Al-Dhahir, ``Reconfigurable intelligent surfaces: Principles and opportunities,'' \emph{IEEE Commun. Surv. Tutorials}, vol.~23, no.~3, pp. 1546--1577, third quarter 2021.

\bibitem{ref7}
Z.~Chen, G.~Chen, J.~Tang, S.~Zhang, D.~K. So, O.~A. Dobre, K.~K. Wong, and J.~Chambers, ``Reconfigurable-intelligent-surface-assisted {B}5{G}/6{G} wireless communications: Challenges, solution, and future opportunities,'' \emph{IEEE Commun. Mag.}, vol.~61, no.~1, pp. 16--22, Jan. 2023.

\bibitem{ref8}
J.~An, C.~Xu, L.~Gan, and L.~Hanzo, ``Low-complexity channel estimation and passive beamforming for {RIS}-assisted {MIMO} systems relying on discrete phase shifts,'' \emph{IEEE Trans. Commun.}, vol.~70, no.~2, pp. 1245--1260, 2022.

\bibitem{ref9}
J.~An, C.~Yuen, C.~Xu, H.~Li, D.~W.~K. Ng, M.~Di~Renzo, M.~Debbah, and L.~Hanzo, ``Stacked intelligent metasurface-aided {MIMO} transceiver design,'' \emph{IEEE Wirel. Commun.}, vol.~31, no.~4, pp. 123--131, 2024.

\bibitem{ref10}
J.~An, C.~Yuen, M.~D. Renzo, M.~Debbah, H.~Vincent~Poor, and L.~Hanzo, ``Flexible intelligent metasurfaces for downlink multiuser {MISO} communications,'' \emph{IEEE Trans. Wirel. Commun.}, pp. 1--1, 2025.

\bibitem{ref11}
Z.~Yu, J.~An, E.~Basar, L.~Gan, and C.~Yuen, ``Environment-aware codebook design for {RIS}-assisted {MU-MISO} communications: Implementation and performance analysis,'' \emph{IEEE Trans. Commun.}, vol.~72, no.~12, pp. 7466--7479, 2024.

\bibitem{ref12}
C.~Liu \emph{et~al.}, ``Coverage probability analysis of {RIS}-assisted high-speed train communications,'' in \emph{Proc. IEEE WCNC}, Glasgow, United Kingdom, Mar 2023, pp. 1--6.

\bibitem{ref13}
A.~Habib, A.~E. Falou, C.~Langlais, and M.~Berbineau, ``Reconfigurable intelligent surface assisted railway communications: A survey,'' in \emph{Proc. IEEE VTC-Spring}, Florence, Italy, Jun 2023, pp. 1--5.

\bibitem{ref14}
J.~Zhang, H.~Liu, Q.~Wu, Y.~Jin, Y.~Chen, B.~Ai, S.~Jin, and T.~J. Cui, ``{RIS}-aided next-generation high-speed train communications: Challenges, solutions, and future directions,'' \emph{IEEE Wirel. Commun.}, vol.~28, no.~6, pp. 145--151, Dec 2021.

\end{thebibliography}
\end{document}